# Nonlinear Dynamics Semi-classical Model of Quantum Spin

*Joshua J. Heiner*
*Department of Physics and Astronomy*
*University of Wyoming*
*1000 E University Ave*
*Laramie, WY 82071*
*jheiner2@uwyo.edu*

*David R. Thayer*
*Department of Physics and Astronomy*
*University of Wyoming*
*1000 E University Ave*
*Laramie, WY 82071*
*drthayer@uwyo.edu*

A nonlinear dynamics semi-classical model is used to show that standard quantum spin analysis can be obtained. The model includes a classically driven nonlinear differential equation with dissipation and a semi-classical interpretation of the torque on a spin magnetic moment in the presence of a realistic magnetic field, which will represent two equilibrium positions. The highly complicated driven nonlinear dissipative semi-classical model is used to introduce chaos, which is necessary to produce the correct statistical quantum results. The resemblance between this semi-classical spin model and the thoroughly studied classical driven-damped nonlinear pendulum are shown and discussed.

Quantum mechanics; quantum statistics; nonlinear dynamics; deterministic chaos; spin interpretation

## I. INTRODUCTION

In a recent publication it was discussed that it may be possible to understand the quantum mechanical (QM) spin states in a similar method used in deterministic chaos, specifically in driven nonlinear dissipative systems [1]. To clarify, it should be noted that it is well known that it is not possible to construct a precise deterministic model of spin (a local hidden variable model), as dictated by the Bell inequality results [2], associated with a singlet state (a spin zero) pair of spin 1/2 particles. Nevertheless, in that recent publication, it was demonstrated that a local spin model, with an appropriate statistical construct, could reproduce the spin correlation results found in standard quantum mechanics. It is with this insight that numerical simulations using a semi-classical spin model are explored, which incorporates deterministic chaos results to model the needed quantum statistics. Furthermore, in an effort to provide better quantum spin interpretation, as well as to potentially provide a semi-classical model, which could be used to explore a deeper level of quantum mechanics, we offer this analysis of deterministic chaos that is shown to agree with the probabilistic spin results found in a Stern-Gerlach device.

At this point and throughout this paper we point out the similarities between the semi-classical spin model that will be presented and the well-known classical driven-damped nonlinear pendulum, which exhibits chaos [3-6].

For simplicity, the geometry used to describe the semi-classical model of a spin 1/2 particle, which is allowed to pass through a Stern-Gerlach spin detector experiment, is a 2D $(x, z)$ rectangular coordinate system (which is sufficiently general to demonstrate our simulation result, although the model can easily be extended to a 3D simulation if desired).

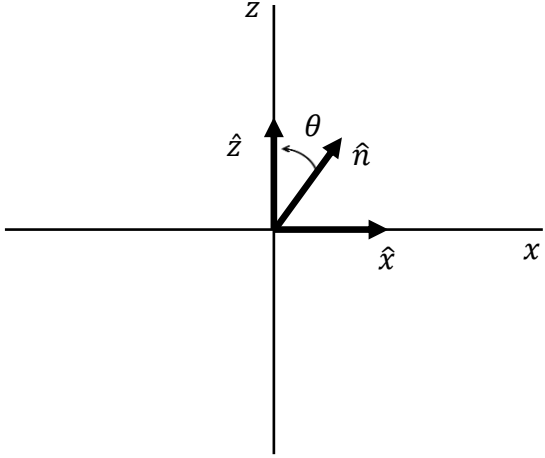

Figure 1: 2D Geometry for the Semi-Classical Spin Model

The dominant magnetic field, $B$, gradient, $\partial B_z/\partial z$, is in the unit vector $\hat{z}$ direction, while the dominant particle propagation is in the perpendicular unit vector $\hat{x}$ direction, as it passes through the spin detector. In order to observe a spin up state, $|\alpha_z\rangle$, versus a spin down state, $|\beta_z\rangle$, associated with the $\hat{z}$ direction, a particle trajectory is found to deviate in the positive versus the negative $\hat{z}$ direction, respectively, away from the primary $x$ axis direction. As is well known using standard quantum mechanical spin analysis [7], any spin state, $|\psi\rangle$, that enters the device can be expressed as a linear combination of the up and down states associated with the $z$ spin basis, as:

$$|\psi\rangle = a|\alpha_z\rangle + b|\beta_z\rangle, \quad (1)$$

where $a$ and $b$ are complex numbers.

After the particle passes through the detector, where the original state has effectively collapsed to $|\psi_{\text{collapsed}}\rangle$, it will either be in a spin up (+) state, as $|\psi_{\text{collapsed}}\rangle = |\alpha_z\rangle$, or a spin down (−) state, as $|\psi_{\text{collapsed}}\rangle = |\beta_z\rangle$, associated with the $z$ spin basis, where the probabilities of occurrence, $P_+$ or $P_-$, are respectively given by:

$$\begin{aligned} P_+ &= |\langle \alpha_z|\psi\rangle|^2 = |a|^2 \\ P_- &= |\langle \beta_z|\psi\rangle|^2 = |b|^2. \\ P_+ + P_- &= |a|^2 + |b|^2 = 1 \end{aligned} \quad (2)$$

Furthermore, for a general spin up and down basis, $|\alpha_n\rangle$ and $|\beta_n\rangle$, associated with the unit vector $\hat{n}$ direction, that makes an angle $\theta$ away from the $\hat{z}$ direction (figure 1), where

$$\hat{n} = \hat{z}\cos\theta + \hat{x}\sin\theta, \quad (3)$$

it is also well known that the connection between both of these spin bases is constructed as:

$$\begin{aligned} |\alpha_n\rangle &= \cos(\theta/2)|\alpha_z\rangle + \sin(\theta/2)|\beta_z\rangle \\ |\beta_n\rangle &= \sin(\theta/2)|\alpha_z\rangle - \cos(\theta/2)|\beta_z\rangle \end{aligned}. \quad (4)$$

Consequently, if an original state is known to be spin up along the $\hat{n}$ direction, as $|\alpha_n\rangle$, then the probability to realize a collapsed spin up, $|\psi_{\text{collapsed}}\rangle = |\alpha_z\rangle$, or a spin down, $|\psi_{\text{collapsed}}\rangle = |\beta_z\rangle$, state associated with the $\hat{z}$ direction are:

$$\begin{aligned} P_+ &= |\langle \alpha_z|\alpha_n\rangle|^2 = \cos^2(\theta/2) \\ P_- &= |\langle \beta_z|\alpha_n\rangle|^2 = \sin^2(\theta/2) \end{aligned}. \quad (5)$$

In a Stern-Gerlach device the classical force that acts on the spin magnetic moment magnitude, $\mu$, once it has collapsed or locked into an up or down spin state is

$$F = \pm\mu\vec{\nabla}B. \quad (6)$$

Here it should be noted that quantum mechanical spin is always parallel to the magnetic field, which results in the observed up or down trajectories found in the Stern-Gerlach experiment.

In the case of the semi-classical spin model, the spin oscillates during a very short time scale as it enters the simulated spin detector (the cause of the oscillation will be discussed later). Ultimately, the spin quickly settles to become parallel to the dominant $\hat{z}$ directed magnetic field, $B$. In a similar fashion as is found for spin $1/2$ particles in a Stern-Gerlach experiment, the semi-classical spin model also dictates precisely two particle trajectories (due to spin up or down states).

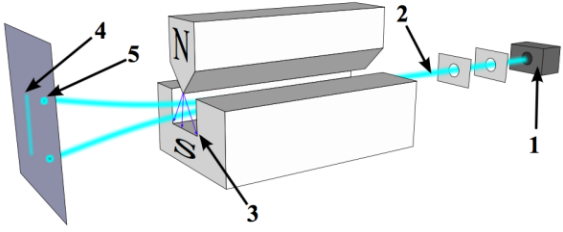

Figure 2: The Stern-Gerlach device and the experimental outcome [8]. The labels are as follows: 1-Furnace, 2-Beam of silver atoms, 3-Inhomogeneous magnetic field, 4-Classical prediction, and 5-Quantum observance.

## II. SEMI-CLASSICAL SPIN

### A. Introduction

The original Stern-Gerlach experiment, as seen in figure 2, sent silver atoms through a non-uniform magnetic field; particularly, a field that exhibits a dominant $\partial B_z/\partial z$. The atoms are sent in along the $x$ axis and experience a force along the $z$ axis due to equation (6).

The classical expectation of a magnetic moment trajectory exhibits a continuous spread in detector positions. However, it is well known that a quantum mechanical spin magnetic moment exhibits only two possible trajectories (due to the spin being in an up or down state).

### B. Classical Magnetic Moment

The torque that a classical magnetic moment, $\vec{\mu}_c$, would feel in the presence of a magnetic field, is

$$I\ddot{\theta} = -b\dot{\theta} - \mu_c B \sin\theta, \qquad (7)$$

where $I$ is the moment of inertia, and $b$ is a factor in the dissipative force.

Due to the torque of the magnetic moment with the magnetic field in equation (7), the classical magnetic moment will reach a stable equilibrium when it is pointing in the same direction as the magnetic field, at $\theta = 0$.

### C. Semi-Classical Magnetic Moment

In order to properly model the torque acting on the magnetic moment so that there are two equilibrium spin positions (up, at $\theta = 0$, and down at, $\theta = \pi/2$), the semi-classical spin model incorporates a two-side torque function of angle, $\theta$. Instead of using the sine function appropriate for a classical torque, the semi-classical torque model is described in the following. It is practical to describe the torque for alignment with the magnetic field as:

$$\tau_+ = -\mu B_z \sin^2 2\theta, \qquad (8)$$

in the bounds,

$$0 \leq \theta \leq \pi/2. \qquad (9)$$

Furthermore, for the case where the spin is anti-aligned with the magnetic field the torque is:

$$\tau_- = \mu B_z \sin^2 2\theta, \qquad (10)$$

in the bounds,

$$\pi/2 \leq \theta \leq \pi. \qquad (11)$$

For consistency in the torque functions, it should be noted that they are each anti-symmetric about their respective equilibrium positions, that is, for equation (8) around $\theta = 0$, and for equation (10) around $\theta = \pi$. For simplicity of presentation, the torque model and differential equation for angle, $\theta(t)$, is shown just for the first case, equation (8) in the bounds set by condition (9).

As seen in figure 3, the torque is zero at $\pi/2$. In comparison to the classical torque at $\pi/2$, shown in the inset, it is at maximum

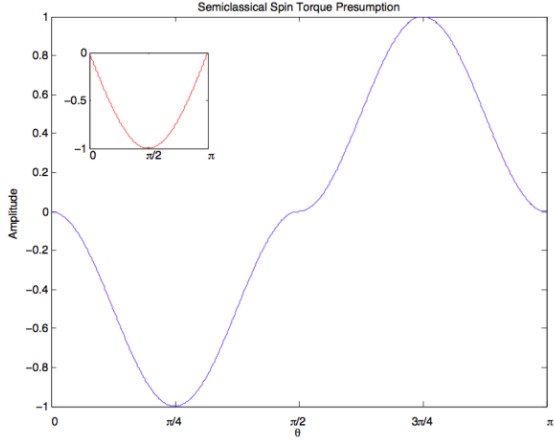

Figure 3: The spin torque is represented in blue from equations and conditions (8)-(11). A classical torque is represented in the inset to show that it is not a realistic representation of the spin torque.

torque magnitude, which is unrealistic for a quantum spin model.

The semi-classical spin model incorporates a realistic magnetic field vector, with $B_x$ and $B_z$ components (as shown in figure 4). The modeled net torque for a quantum spin is:

$$I\ddot{\theta} = -b\dot{\theta} - \mu B_z \sin^2 2\theta \\ + \mu B_x \sin^2 2\theta \quad . \quad (12)$$

The negative sign in front of the $B_z$ term tends to lead to an equilibrium magnetic moment position at $\theta = 0$, while the positive sign in front of the $B_x$ term tends to lead to an equilibrium magnetic moment position at $\theta = \pi/2$.

Letting $\dot{\theta} = y$, and representing the magnetic field as changing in time, due to the independent particle trajectory through the magnetic field, the system described in equation (12) becomes:

$$\dot{\theta} = y, \quad (13)$$

$$\dot{y} = -I^{-1}(by + \mu B_z(t) * \sin^2 2\theta \\ -\mu B_x(t) * \sin^2 2\theta) , (14)$$

which is strikingly similar to the governing equations for a classical driven-damped nonlinear pendulum:

$$\dot{\theta}' = y', \quad (15)$$

$$\dot{y}' = -a'y' - b'\sin\theta + c'F(t). \quad (16)$$

The constants $a'$, $b'$, and $c'$ are well known parameters and $F(t)$ is a driving force (normally sinusoidal) [9].

The Poincare-Bendixson theorem concludes that chaos cannot occur in a two-dimensional phase plane [10-11]. It can be shown; however, that for both cases the system is non-autonomous, meaning it can be represented using a higher order autonomous system for which chaos can occur [12-13]. It is worth noting that chaos in a pendulum can only occur when the drive force is greater than the force due to gravity, $c'F(t)/b' > 1$. Ultimately, this semi-classical spin model also exhibits chaos due to similarly sufficient conditions being satisfied.

### D. Real Magnetic Field

It is important to note that although the standard Stern-Gerlach experiment focuses primarily on a gradient of the magnetic field in the dominant $\hat{z}$ direction, since $\nabla \cdot B = 0$, there must also exist a gradient and magnetic field in the $\hat{x}$ direction.

Since the spin is highly sensitive to a magnetic field, it is important to represent a real magnetic field before the spin enters the Stern-Gerlach device.

In order to observe chaotic results, which are necessary to obtain the expected quantum spin statistics, the $B_x$ torque term in equation (14) has to be larger than the $B_z$ term for a small period of time/space, which is realistically valid as the spin enters the magnetic field towards the origin as shown in figure 4. The condition of observable chaos is once again similar to the driven-damped nonlinear pendulum, $B_x/B_z > 1$, where $B_x$

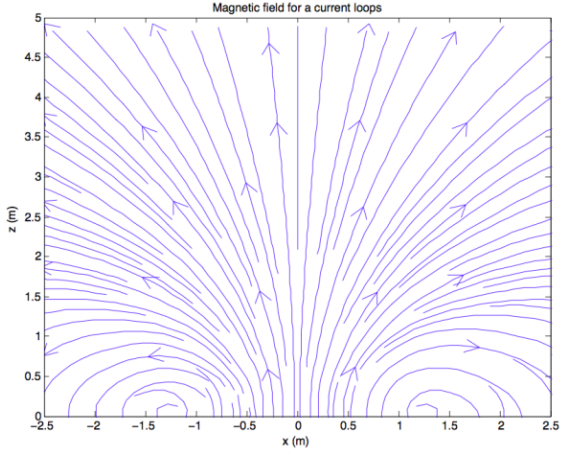

Figure 4: The magnetic field of a current loop is shown with the center position at (0,0).

can be thought of as the independent driving force leading to chaos.

For example, as shown in figure 4 the magnetic field at $z = 0.05m$ goes from negative to positive $B_z$, as the particle moves along $x$. During that transition the $B_x$ term will be dominant for a short period of time/space, thus providing the needed driving force to obtain chaotic results.

For the numerical simulation, the magnetic field will be represented as one current loop since it describes a similar magnetic field of the Stern-Gerlach device.

### E. Code-Matlab

The numerical method used to step through equations (14) and (13) is the forward Euler-Cromer method. If the standard Euler method was used it would result in an increase in energy, making it less accurate and not realistic.

In standard quantum mechanics a time scale uncertainty, $\Delta\tau_{QM}$, is often represented in terms of a lower or minimum limit:

$$\Delta\tau_{QM} \geq \frac{\hbar}{2\Delta E}, \qquad (17)$$

where $\Delta E$ is the energy scale uncertainty.

This can also be thought of as the minimum time required to measure an energy scale uncertainty, $\Delta E$, in the quantum domain. In this Stern-Gerlach model the energy difference between the up and down spin state is the energy uncertainty, $\Delta E = 2\mu B$.

Ultimately, the time steps, $\Delta t$, in the numerical simulation needs to be much less than the standard quantum mechanics time scale:

$$\Delta t \ll \Delta\tau_{QM}. \qquad (18)$$

Due to computational limits, $\Delta t$ is roughly three orders of magnitude smaller than $\Delta\tau_{QM}$.

In an attempt to recreate the Stern-Gerlach experiment, the velocity of the particle will be constant at 500m/s.

Standard QM spin statistics describes what is observed and what a credible model should obtain. From equation (5), the spin down probability statistic, $P_-$, will be used to compare to the numerical simulation.

### III. RESULTS

Initial spin orientation values, $\theta_i$, were divided equally from 0 to $\pi$ in increments of $\pi/1001$. An odd number is always used to prevent an initial state of exactly $\pi/2$, which would never numerically flip to either spin up or down states.

Figure 5 shows the numerical simulation results of equations (14) and (13). Note that the actual simulation data results are always zero or one (corresponding to the spin down or up state, respectively). The data plotted in figure 5 is a combination of two averages. First, the simulation is run 100 times and the results for each initial angle, $\theta_i$, are averaged, i.e. if the result for one $\theta_i$ provides spin up 10 times and spin down 90 times, then the resultant average probability is 0.1. The second is an average of the nearest neighbors in angle space (described in the following section), i.e. if just the two nearest neighbors around the angle of $\theta_i$ are considered; the angles of interest are $\theta_i \pm \pi/1001$. Between each simulation the time step is slightly changed and the initial position is also changed in the third significant figure.

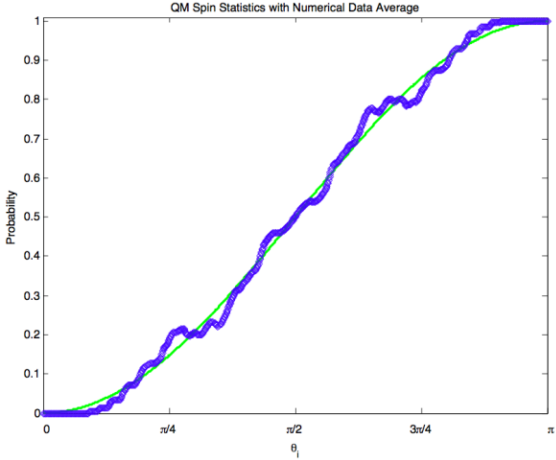

Figure 5: The green line is the QM Spin Statistic of a particle being spin down. The blue points are the averaged simulation spins comprised of 1001 different initial angles ranging from 0 to $\pi$. The simulation was done 100 times with slightly different initial conditions.

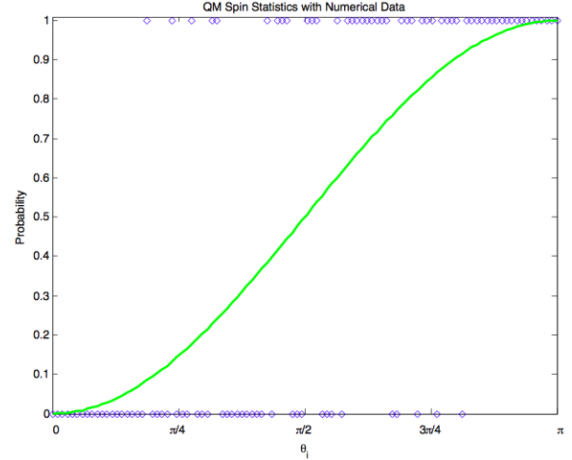

Figure 6: The green line is the QM Spin Statistic of a particle being spin down. The blue numerical data points are obtained from equations (14) and (13). This simulation only had 101 initial conditions, so the reader could observe what is happening, and only shows one simulation run. This figure is only for the reader to understand the result of one simulation run.

Since the system exhibits chaos, both the time step and initial position need to slightly change for each run to eliminate the effects of numerical attractors and fractal dimension.

One simulation run, using increments of $\pi/101$, is shown only for the benefit of the reader in figure 6. As noted earlier, the outcome of a single simulation results in spins that are always zero or one (corresponding to the spin down or up state, respectively). There appears to be an attractor at $\theta_i \sim 3\pi/8$. At this location, there appears to be a clump of data points that are attracted to being in the spin down state. Changing the parameters and averaging over nearest neighbors in angle space slightly helps to eliminate most non-realistic attractors, but sometimes they still appear due to the discrete nature of numerical simulations. The fractal nature can be observed from figure 5; it appears that the averaged data changes in stair steps that have a similar width (this will be discussed later).

The parameters $b$ and $I$ were adjusted accordingly to produce the results that would match up with the statistical quantum mechanical results. If the initial position, velocity, or magnetic field changes by too much, the results obtained change dramatically and do not match with statistics. It should be noted; however, that a slight adjustment in the parameters $b$ and $I$ bring the data back to matching with the expected statistics. This shows robustness in the model.

Finally, the simulation data represented in figure 5 is obtained by taking the raw data of one hundred simulated runs, where ultimately nearest neighbor in angle space results are averaged using a convolution. The convolution average was obtained using 150 nearest neighbors in angle space (75 on each side), i.e. a square wave of magnitude one and length 150. The end values at 0 and $\pi$ were extended with values of zeros and ones to give a realistic representation of the boundary conditions.

The individual spin trajectories are shown in figure 7. The spin is highly sensitive to the initial condition which is demonstrated from $\theta_i = 313\pi/1001$ to $\theta_i = 314\pi/1001$. A small change in the initial conditions of $\pi/1001$ led to a completely opposite outcome.

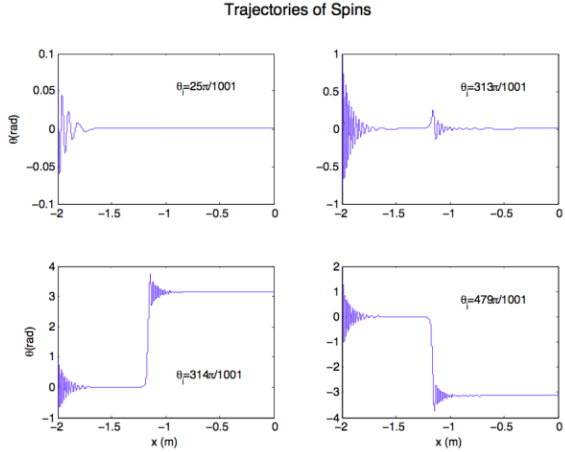

Figure 7: Four different initial conditions of $\theta$ are plotted as trajectories. The top right and bottom left plots differ by an initial condition of $\pi/1001$, showing the sensitivity to initial conditions.

## IV. CHAOS ANALYSIS

Although there are many mathematical methods to analyze chaotic systems [14], the purpose of this paper is not to thoroughly investigate the observed chaos (although this could prove to be equally interesting), but to show that quantum statistics can be reproduced using a driven nonlinear dissipative semi-classical model. We do however show a crude analysis of the chaos structure in order to briefly show that the results need to be smoothed to a dimension near one to compare with the statistical quantum mechanics.

Due to the observed sensitivity to initial conditions and the fractal nature of the results, the driven nonlinear dissipative equation (12) should be described as leading to deterministic chaos. Below, the fractal nature of the chaos is briefly explored.

Chaotic numerical systems often exhibit trajectories or paths that converge to attractors. As a result, it is informative to explore the fractal dimension of the attractor. Using this approach to quantify the degrees of freedom in the system helps to elucidate the systems complexity.

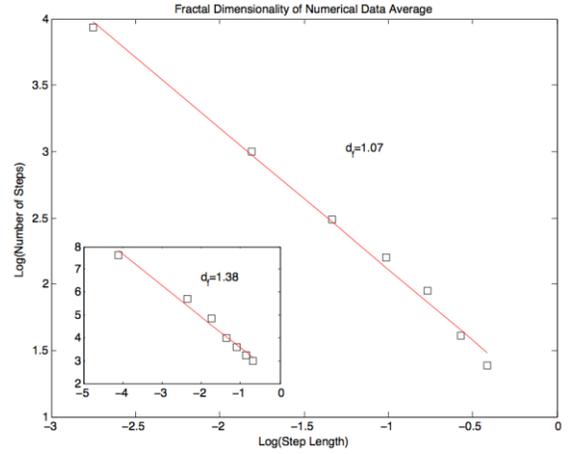

Figure 8: The fractal dimensionality of the numerical data average for 150 nearest neighbors in angle space is shown. The inset is the fractional dimensionality of the raw data set before averaging nearest neighbors in angle space.

The stair steps of seemingly equal width that were mentioned from the numerical data in figure 5, which have been smoothed to a dimension near one (noted later as being $d_f = 1.07$), exhibit fractal nature. Giordano and Nakanishi have presented a numerical procedure to analyze the fractal dimensionalities of curves [15], which was originally discussed by Kolmogorov and others [16-19]. This method compares how many steps are needed, $k$, to measure the length of the curve while changing the step length, $L(k)$. The number of steps varies as

$$k * L(k) \sim L(k)^{1-d_f}, \quad (19)$$

where $d_f$ is the fractal dimension and can be solved as

$$d_f = -\frac{\ln k}{\ln L(k)}. \quad (20)$$

Figure 8 shows the result of equation (20) for the averaged data shown in figure 5. The same process was performed before the averaging and also for the averaging of just 50 nearest neighbors in angle space ($nas$), where the fractal dimensions are:

$$d_{f\_raw} = 1.38$$
$$d_{f\_50nas} = 1.12 \quad . \quad (21)$$
$$d_{f\_150nas} = 1.07$$

The fractal dimension is extremely prevalent in the raw data, such that, as the averaging increases it becomes closer to 1D. The averaging did not span more than $150 nas$ because it becomes less realistic to average over a large quantity of $nas$. Ultimately, it is important to note that the simulation results (shown in figure 5) were presented using a convolution averaging parameter of $nas = 150$, since the resultant curve, with $d_{f\_150nas} = 1.07$, is very close to 1D, which is important to use when comparing the simulation results to the 1D QM spin statistic curve.

## V. CONCLUSION

The results obtained from the semi-classical model that has been presented could result into a deeper insight into the complexity of the quantum mechanical spin states.

In using a deterministic chaotic dynamics model as suggested in a recent publication, [1], which should not be confused with the failure of precise determinism as stated by the Bell inequality analysis [2], the results still exhibit the usual unpredictability or uncertainty that is prevalent in standard quantum mechanics.

Having obtained similar results to quantum mechanical spin statistics using the semi-classical spin model, we look forward to further analysis using more realistic magnetic fields associated with actual Stern-Gerlach devices. Specifically, further understanding may be gained using a full 3D model of the magnetic field, as well as the semi-classical trajectory of the spin particle.

## ACKNOWLEDGEMENTS

We are thankful for the generous support from the Department of Physics and Astronomy at the University of Wyoming.

[1] Thayer, D. R., and Jafari, F., *Int. J. Ad. Res. Phys. Sc.* **2** 18-26 (2015)
[2] Bell, J. S., *Physics* **1** 195 (1964)
[3] Markus, L., "Lectures in differentiable dynamics", *Amer. Math. Soc.*, Providence, R.I., (1971)
[4] Holmes, P., *Appl. Math. Modeling* **1** 362-366 (1977)
[5] D'Humieres, D., Beasley, M. R., Huberman, B. A., Libchaber, A., *Phys. Rev. A* **26** 3483-3496 (1982)
[6] Kerr, W. C., Williams, M. B., Bishop, A.R., et al. *Phys. B – Condensed Matter* **59** 103-110 (1985)
[7] McIntyre, D. H., "Quantum Mechanics: A Paradigms Approach", Pearson Education, Inc. (2012).
[8] Tatoute - Own work, CC BY-SA 4.0, https://commons.wikimedia.org/w/index.php?curid=34095239
[9] Thornton, S. T., Marion, J. B., "Classical Dynamics of Particles and Systems" 5th Edition, Brooks/Cole (2008)
[10] Poincare H., *J. de Math.* **7** 375-422 (1881); **8**, 251-296 (1882); *Ann. Math.* **8** 553-564 (1882)
[11] Bendixson, I., *Acta Math.* **24** 1-88 (1901)
[12] Varghese, M., Thorp, J. S., *Phys. Rev. Lett.* **60** 665-668 (1988)
[13] Bevivino, J., *Dynamics at the Horseshoe* **1** 1-24 (2009)
[14] May, R. M., *Nature* **261** 459-467 (1976)
[15] Giordano, N. J., Nakanishi, H., "Computational Physics" 2nd Edition, Addison-Wesley (2006)
[16] Kolmogorov, A., "On Tables of Random Numbers" *Sankhya Ser. A.* **25** 369-375 (1963)
[17] Mandelbrot, B., *Science* **156** 636-638 (1967)
[18] Mandelbrot, B., "The Fractal Geometry of Nature" Rev. Edition, W. H. Freeman and Company (1982)
[19] Higughi, T., *Physica D* **31** 277-283 (1988)